\input ./style/arxiv-general.cfg
\documentclass[aoas,MSNbibl,nameyear,dvips]{arximspdf}
\makeatletter
   \@ifpackageloaded{graphicx}{}{\usepackage{graphicx}}
\makeatother
\usepackage{url,breakurl}

\doi{10.1214/15-AOAS839}
\volume{9}
\issue{3}
\pubyear{2015}
\firstpage{1169}
\lastpage{1193}
\docsubty{FLA}

\makeatletter

\newcommand{\eqref}[1]{(\ref{#1})}
\newcommand{\bl}[1]{{\mathbf #1}}
\newcommand{\bs}[1]{{\bolds #1}}

\newcommand{\tr}{\operatorname{tr}}
\newcommand{\Exp}[1]{{\mathrm{E}}[ #1 ]  }
\newcommand{\Cov}[1]{{\operatorname{Cov}}[  #1 ]  }
\newtheorem{prop}{Proposition}
\makeatother

\begin{document}
\begin{frontmatter}

\title{Multilinear tensor regression for longitudinal relational data}
\runtitle{Multilinear tensor regression}

\begin{aug}
\author[A]{\fnms{Peter D.}~\snm{Hoff}\corref{}\ead[label=e1]{pdhoff@uw.edu}\thanksref{T1}}
\runauthor{P.~D. Hoff}
\thankstext{T1}{Supported by NIH Grant R01HD067509.}

\affiliation{University of Washington}
\address[A]{Departments of Statistics and Biostatistics\\
University of Washington\\
Seattle, Washington 98195-4322\\
USA\\
\printead{e1}}
\end{aug}

%
\received{\smonth{11} \syear{2014}}
%
\revised{\smonth{5} \syear{2015}}

%
\begin{abstract}
A fundamental aspect of relational data, such as from a social
network, is the possibility of dependence among the relations.
In particular, the relations between members of one pair
of nodes may have an effect on the relations between members of
another pair. This article develops a type of regression
model to estimate such effects in the context of longitudinal and
multivariate relational data, or other data
that can be represented in the form of a tensor.
The model is based on a general multilinear tensor regression model,
a~special case of which is
a tensor autoregression model
in which
the tensor of relations at one time point are
parsimoniously
regressed on relations from previous time points.
This is done via a separable, or Kronecker-structured,
regression parameter along with a separable
covariance model.
In the context of an analysis of longitudinal multivariate
relational data, it is shown how
the multilinear tensor regression model
can represent
patterns
that often appear in relational and network data,
such as reciprocity and transitivity.
\end{abstract}

%
\begin{keyword}
\kwd{Array normal}
\kwd{Bayesian inference}
\kwd{event data}
\kwd{international relations}
\kwd{network}
\kwd{Tucker product}
\kwd{vector autoregression}
\end{keyword}
\end{frontmatter}

\section{Introduction}\label{sec1}

Longitudinal relational data among a set of $m$ objects or nodes
can be represented as a time series
of matrices $\{ \bl Y_t: t=1,\ldots, n\} $, where
each $\bl Y_t $ is an $m\times m$
square matrix. 
The entries of $\bl Y_t$
represent directed relationships or actions involving
pairs of nodes (dyads) at time $t$, so $y_{i_1,i_2,t}$ is
a numerical description of the
action taken by node $i_1$ with node $i_2$ as the target at time $t$.
Such data therefore consist of a time series for
each pair of nodes.
For example, in this article we consider longitudinal data on
actions involving country pairs, where
$y_{i_1,i_2,t}$ represents the intensity of actions
taken by country $i_1$ toward country $i_2$ in time period $t$.
Specifically, we analyze
weekly relational measures between pairs of 25 countries
over the roughly ten and a half year period from 2004 to mid-2014,
giving $n=543$ weeks of data.
The value of $y_{i_1,i_2,t}$ is a transformed count of the number of positive
verbal statements of country $i_1$ toward country $i_2$ during week~$t$
(a fuller description of the data appears in Section~\ref{sec4}).

\begin{figure}

\includegraphics{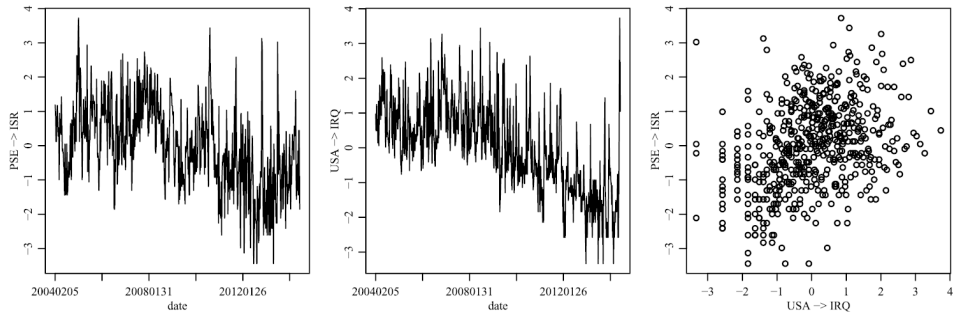}

\caption{From left to right, positive verbal relations versus time from
Palestine to Israel and
USA to Iraq, and a scatterplot.}\vspace*{-3pt}
\label{fig:exdata}
\end{figure}

While the statistical challenge in analyzing static relational data
is to describe the potential dependence between dyadic observations,
with longitudinal data the challenge
is to describe dependence between dyadic time series.
Such dependence in our data set is illustrated graphically in
Figure~\ref{fig:exdata}: Two dyadic
time series are positively correlated,\vadjust{\goodbreak} even though
they have no nodes in common.
In this article we develop a parsimonious approach
to analyzing and describing such dependencies between time series.
This is done in the context of a statistical model for
the time series of matrices
$\{ \bl Y_t: t=1,\ldots, n\}$. 

%

Foundational development of a class of agent-based
longitudinal network models
appears in
\citet{snijders2001} and is developed further in
\citet{snijderssteglichschweinberger2007}.
These articles develop models for binary relational data (i.e.,
social networks) in which
social links are modeled as the result
of decisions made by nodes acting to maximize their individual utilities.
Parameters in the models
can be interpreted as preferences for various
types of social structures, such as
reciprocated dyads or transitive triads.
These parameters are typically homogeneous, in that
they are common to all individuals in the network
(or possibly common to all individuals having common observable attributes).
Further development of homogeneous models for binary data
has involved the use of exponentially parameterized random graph
models [\citet{hannekefuxing2010,krivitskyhandcock2014}].

A popular alternative to such homogeneous models
utilizes a dynamic latent variable formulation,
in which each $\bl Y_t$ is represented as a function of node-specific latent
variables $\bl Z_t$ that evolve over time. \citet{wardhoff2007},
\citet{wardahlquistrozenas2013} and \citet{durantedunson2014}
model the
relationship between nodes $i_1$ and $i_2$ at time $t$ as a function
of low-dimensional latent variables
$\bl z_{i_1,t}$ and $\bl z_{i_2,t}$.
\citet{hoff2011a} considers a version of such a model where
the latent variables are parameterized as static latent factors
that are modified
by time-varying weights.
Similar models considered by
\citet{fusongxing2009} and \citet{xingfusong2010} assume the
latent variables are
categorical-valued latent classes.
Latent variable models such as these
can be viewed as a class of random effects models,
and can represent
certain types of dependence often seen in social networks and
relational data
[\citet{hoff2008a}].
Somewhat related to this,
\citet{westveldhoff2011} and \citet{hoff2011b} consider different
covariance models for longitudinal relational data.

A fundamental feature of relational data is the statistical
interdependence among relations,\vadjust{\goodbreak} and
a standard goal of relational data analysis is to
quantify and evaluate this interdependence.
The two modeling approaches discussed in the previous paragraph both
represent certain types of dependencies, but in different ways.
The agent-based approach explicitly models how dyads might affect one
another, but
generally assumes such influences are
homogeneous.
Conversely, the
latent variable approach allows for across-node heterogeneity
in the representation of network behavior, but the
interdependence between relations is not explicitly parameterized,
and the types of dependence that can be represented are
limited by the simple structure of the latent variables.
This article presents a modeling approach that is unlike either
the agent-based or the random effects models, but, like the
former, has an explicit representation of the dependence between
dyads, and, like the latter, allows for
nodal heterogeneity in the model parameters.
The approach is based on a reduced-parameter regression model as follows:
Consider modeling the actions $\bl Y_t$ at time $t$
as a function of their values
$\bl X_t \equiv\bl Y_{t-1}$
at the previous time point.
A conceptually simple model for such data would be a
vector autoregressive (VAR) model.
Letting
$\bl y_t=\operatorname{vec}(\bl Y_t)$ and $\bl x_t=\operatorname
{vec}(\bl X_t)$, a~first-order
VAR model posits that
\[
\bl y_t = \bs\Theta\bl x_t + \bl e_t, \qquad{
\mathrm{E}}[\bl e_t ] = \bl0,\qquad {\mathrm{E}}\bigl[ \bl e_t
\bl e_s^T \bigr] = \cases{ %
\Sigma,& \quad$\mbox{if } t=s,$
\vspace*{2pt}\cr
\bl0, &\quad $\mbox{if } t\neq s,$ } 
\]
where $\bs\Theta$ and $\Sigma$ are parameters to be estimated.
For simplicity,
here and in what follows we consider models without intercepts, which are
appropriate
if the time series for each pair $i_1,i_2$ has been demeaned
(so that $\sum_{t} y_{i_1,i_2,t} /n = 0$).
Given sufficient data,
unrestricted estimates of
$\bs\Theta$ in a VAR model can generally be obtained via
ordinary least squares (OLS) or
feasible generalized least squares (GLS). However, such estimates
can be unstable or unavailable unless the time series is extremely long:
As $\bl y_t$ and $\bl x_t$ are each of length
$m^2$ [or $m(m-1)$ if the diagonal of each $\bl Y_t$ is undefined],
the regression matrix $\bs\Theta$ has $m^4$ entries ($m^2$ per pair of nodes).

Estimation stability can be improved by restricting $\bs\Theta$
to belong to a parameter space of lower dimension.
In this article we focus on models
where the regression matrix has the form
$\bs\Theta= \bl B \otimes\bl A$, where $\bl A$ and $\bl B$
are $m\times m$ matrices and ``$\otimes$'' is the Kronecker product.
Such a model is a ``bilinear'' regression model, as in terms
of $\bl Y_t$ and $\bl X_t$ the model is
%
%
\begin{equation}
\bl Y_t = \bl A \bl X_t \bl B^T + \bl
E_t, \label{eqn:blrt}
\end{equation}
so that the regression model is bilinear in the parameters,
that is, linear in $\bl A$ and linear in $\bl B$, but not linear in
$(\bl A,\bl B)$.
This model appears similar to, but is distinct from, the
``growth curve'' model
[\citet{potthoffroy1964,gabriel1998,srivastavarosenrosen2009}], in which
$\Exp{\bl Y|\bl X, \bl Z, \bl C} = \bl X \bl C \bl Z^T$,
where $\bl X$ and $\bl Z$ are known and $\bl C$ is
a matrix of parameters to be estimated. This latter model is linear in the
parameters and bilinear in the two explanatory matrices $\bl X$ and
$\bl Z$.
The model in (\ref{eqn:blrt}) is more related to recently developed
reduced-rank regression
models
[\citet{basudunaganduhmuniswamy-reddy2012},
\citet{shixubaraniuk2014},
Li, Zhou and Li (\citeyear{lizhouli2013})],
in which a scalar response $y$
is regressed on a matrix
$\bl X$ via the mean function $\tr(\bl C\bl X \bl D^T)$,
where
$\bl C \in\mathbb R^{r_1\times p_1}$ and $\bl D \in\mathbb
R^{r_2\times p_2}$,
with $r_1<p_1$ and $r_2<p_2$.
In particular, a~rank-one model has the mean function
$ \bl c^T \bl X \bl d$, with $\bl c \in\mathbb R^{p_1}$ and
$\bl d \in\mathbb R^{p_2}$. Similarly,
in model (\ref{eqn:blrt})
the mean function for
element $i_1,i_2$ of $\bl Y_t$ is given by
$\bl a_{i_1}^T \bl X_t \bl b_{i_2}$, and
so (\ref{eqn:blrt}) can
be seen as a rank-one regression model for each dyad $i_1,i_2$, but
one in which the parameters are shared across dyads.
This parameter sharing leads to $m$-times fewer parameters
than having separate rank-one models for each dyad
(roughly $2m^2$ versus $2m^3$ parameters).
This reduction in the number of parameters, in addition to the
information sharing across dyads that it allows,
can be helpful when the amount of data is limited. For example,
as will be shown in an example data analysis,
using separate rank-one regression models for each dyad can lead to
severe overfitting as compared to model (\ref{eqn:blrt}).


Interpretation of the parameters in (\ref{eqn:blrt}) is facilitated by
noting that
for a given ordered pair of nodes $(i_1,i_2)$,
\[
{\mathrm{E}}[ y_{i_1,i_2,t} | \bl X_t ] = \sum
_{j_1} \sum_{j_2 } a_{i_1,j_1}
b_{i_2,j_2} x_{j_1,j_2}.
\]
Roughly speaking,
$a_{i_1,j_1}$ describes how
the actions by $i_1$ are
influenced by previous actions of $j_1$, and
$b_{i_2,j_2}$ describes how actions toward
$i_2$ are influenced by previous actions toward $j_2$.
This model could be referred to as a multiplicative model,
as the element of the regression coefficient matrix $\bs\Theta$
corresponding to $(y_{i_1,i_2}, x_{j_1,j_2})$ is given by
$a_{i_1,j_1} b_{i_2,j_2}$, and so
is a multiplicative function of
the parameters. 
A more familiar analogue to this multiplicative model is an
additive model such as
\begin{eqnarray*}
y_{i_1,i_2,t}& =& \sum_{j_1} \sum
_{j_2 } (a_{i_1,j_1} + b_{i_2,j_2} ) x_{j_1,j_2}
+\varepsilon_{i_1,i_2,t},
\\
\bl Y_t & = &\bl A \bl X_t \bl1 \bl1^T +
\bl1 \bl1^T \bl X_t \bl B^T + \bl
E_t,
\end{eqnarray*}
where ``$\bl1$'' denotes a vector of $m$ ones.
While
perhaps in an unfamiliar form, this additive model can be expressed as an
ordinary linear regression model, although with a complicated design matrix.
The additive and multiplicative models have essentially the same number of
parameters,
but their interpretation is somewhat different.
In the multiplicative model, the influence of $x_{j_1,j_2}$ on $y_{i_1,i_2}$
is nonnegligible if
\emph{both} $a_{i_1,j_1}$ and $b_{i_2,j_2}$ are nonnegligible. In the additive
mode, $x_{j_1,j_2}$ influences $y_{i_1,i_2}$ if \emph{either}
$a_{i_1,j_1}$ or $b_{i_2,j_2}$ are nonnegligible.
Which model provides a closer approximation to the data-generating process
will depend on the
application. However, we argue that for many longitudinal
relational data sets,
and longitudinal international relations data in particular,
the effect of $x_{j_1,j_2}$ on $y_{i_1,i_2}$ will be small
for most values of $i_1,i_2,j_1,j_2$, and large
only when there is some similarity between both $i_1$ and $j_1$, and
$i_2$ and $j_2$.
For example, if $i_1$ and $j_1$ have an alliance, and $i_2$ and $j_2$
have an alliance,
then the actions of $j_1$ toward $j_2$ may influence future actions of
$i_1$ toward $i_2$, but perhaps not of $i_1$ toward $i'$, a~country
unallied with
$j_2$.


We evaluate this claim empirically with a
brief comparison of the two models.
We fit both of these models to the
country interaction
data using a least squares criterion.
While the models explain only a small fraction of the data variation,
the multiplicative model explains over twice as much:
The $R^2$ coefficients (one minus the ratio of the
residual sum of squares to the total sum of squares) are
$5.8\%$ for the additive model and $13.2\%$ for the multiplicative model.
As the models have the same number of parameters, this suggests we
should favor
the multiplicative model over the additive model.

\begin{figure}

\includegraphics{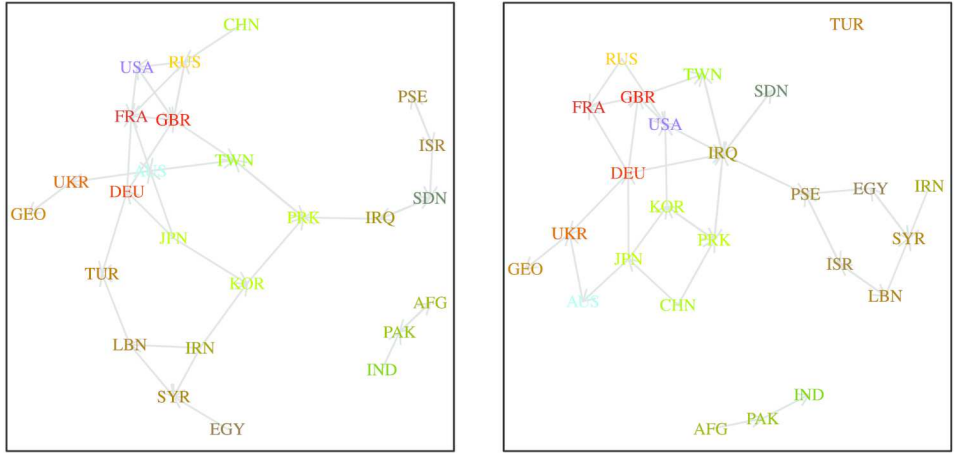}

\caption{Relatively large entries of $\hat{\bl A}$ (left) and $\hat
{\bl B}$ (right).}
\label{fig:bols}
\end{figure}

As there are a large number of parameters in the multiplicative
model (essentially
$2m^2$), it is natural to wonder if they are simply representing
noise in the data or a meaningful signal.\vspace*{1pt} To examine this,
we identified the $i,j$ pairs for which the values of
$\hat a_{i,j}$ and $\hat b_{i,j}$ are largest. This information is
depicted graphically in Figure~\ref{fig:bols}, in which
a link is drawn between countries $i$ and $j$ if
$\hat a_{i,j}$ is among the largest 10\% of values
of $\hat{\bl A}$ (in the left panel) or
$\hat b_{i,j}$ is among the largest 10\% of values
of $\hat{\bl B}$ (the right panel).
The figure indicates a strong geographic component to the
off-diagonal elements of $\hat{\bl A}$ and $\hat{\bl B}$
(plotting labels the standard ISO-3 country codes).
This is empirical
evidence that relations between a pair $(j_1,j_2)$ are in some cases
predictive of future relations between other pairs $(i_1,i_2)$. Otherwise,
these off-diagonal components would be representing noise, and there
would be
no discernible geographic pattern.

We examined this claim further with a small cross-validation study.
We randomly generated 10 cross-validation data sets, each
consisting of
a training set and test set of 488 and 55 values of
$\{\bl Y_t, \bl X_t\}$, respectively.
For each data set, least squares parameter estimates for the additive
and multiplicative models were obtained from each training set,
and then used to
make predictions of each $\bl Y_t$ in the test set.
The multiplicative model outperformed the additive model
for all data sets:
The average predictive $R^2$ for the multiplicative model
was 12.3\% (with a range of 10.9\% to 13.7\%), compared to
4.5\% (with a range of 3.7\% to 5.4\%) for the additive model.

%

Given the modest $R^2$ and predictive $R^2$ values for the multiplicative
model,
it is natural to wonder whether or not a more complex model might achieve
a better fit. For example, one could fit a separate rank-one regression model
for each dyad,
of the form $y_{i_1,i_2,t} =
\bl c_{i_1,i_2}^T\bl X_t \bl d_{i_1,i_2} + \varepsilon_{i_1,i_2,t}$.
Unlike the multiplicative model in (\ref{eqn:blrt}),
the parameters here are distinct for each dyad, and so the number of
parameters is on the order of $2m^3$ instead of $2m^2$ as in the
multiplicative model.
Such an approach does indeed improve within-sample fit, giving an
$R^2$ of 26.5\%. However, applying the cross-validation analysis
to this approach indicates severe overfitting:
The average predictive $R^2$ was $-$2.4\%
(with a range of $-$3.5\% to $-$0.2\%), indicating that using
separate rank-one fits is \emph{worse} than fitting no model, in
terms of identifying consistent patterns in the data.

The performance of the multiplicative model relative
to comparable alternatives
motivates further study and development of 
models of this form. In the next section, we present some basic theory
for this model,
including results on identifiability, convergence of OLS estimates and
parameter interpretation under model
misspecification.
We then extend this model to a general multilinear regression model that
can accommodate longitudinal measurements of multiway arrays, or tensors.
Such models are motivated by the fact that
a more complete version of the data set includes information on four different
relation types, and so the data $\bl Y_t$ at week $t$ consist of a
$25\times25\times4$ three-way tensor.
The regression problem then becomes one of regressing the relational
tensor $\bl Y_t$ from
time $t$ on the tensor $\bl X_t =\bl Y_{t-1}$ from time $t-1$
in a parsimonious way. To accomplish this, in Section~\ref{sec3} we
propose and
develop
the following multilinear generalization
of the bilinear regression model: To relate an $m_1\times\cdots\times
m_K$ tensor $\bl Y_t$ to a $p_1\times\cdots\times p_K$ tensor $\bl
X_t$, we use the model
\begin{eqnarray*}
\bl Y_t & =& \bl X_t \times\{\bl B_1,
\ldots,\bl B_K\} + \bl E_t \quad\mbox{or, equivalently,}
\\
\bl y_{t}&= &( \bl B_K\otimes\cdots\otimes\bl
B_1) \bl x_t + \bl e_t,
\end{eqnarray*}
where ``$\times$'' is a multilinear operator known as the ``Tucker product,''
and
$\bl y_t, \bl x_t, \bl e_t$ are the vectorizations of
$\bl Y_t,\bl X_t, \bl E_t$, respectively.
We present least squares and Bayesian approaches to parameter estimation,
including methods for joint inference on the regression coefficients
and the error variance, $\Cov{\bl e_t} = \Sigma$.
Sample size limitations will generally preclude unconstrained
estimation of $\Sigma$,
an $m\times m$
error covariance matrix,
where $m= \prod m_k$.
As a parsimonious alternative,
we use an array normal model for $\bl e_t$, which is
a multivariate normal model with a Kronecker structured covariance
matrix, $\Cov{\bl e_t} =\Sigma_K \otimes\cdots\otimes\Sigma_1$
[Akdemir and Gupta (\citeyear{akdemirgupta2011}), \citet{hoff2011b}].
Bayesian estimation for the resulting general
multilinear tensor regression model
with Kronecker structured error covariance can be made using semi-conjugate
priors and a Gibbs sampler.

A detailed analysis of the longitudinal relational data presented above
is given in Section~\ref{sec4}. This includes a cross-validation study
to evaluate
different models, development of a parsimonious model
that allows for network reciprocity and transitivity, and a summary of
a Bayesian analysis of the data using this latter model.
A discussion of model limitations and possible extensions follows in
Section~\ref{sec5}.

\section{The bilinear regression model}\label{sec2}
In this section and the next we consider the general problem of
regressing one tensor $\bl Y$ on another tensor $\bl X$, where
$\bl Y$ and $\bl X$ are of potentially different sizes.
We start with the matrix case:
A bilinear regression model of a matrix $\bl Y\in\mathbb R^{m_1\times m_2}$
on a matrix $\bl X \in\mathbb R^{p_1\times p_2}$ takes the form
%
%
\begin{equation}
\bl Y = \bl A \bl X \bl B^T + \bl E, \label{eqn:blr}
\end{equation}
where
$\bl E$ is an $m_1\times m_2$ matrix of mean-zero disturbance terms, and
$\bl A \in\mathbb R^{m_1\times p_1}$ and
$\bl B \in\mathbb R^{m_2\times p_2}$ are unknown matrices to
be estimated. As discussed in the \hyperref[sec1]{Introduction}, this
model can be equivalently represented as
%
%
\begin{equation}
\bl y = (\bl B\otimes\bl A) \bl x + \bl e, \label{eqn:vblr}
\end{equation}
where ``$\otimes$'' is the Kronecker product and
$\bl y$, $\bl x$ and $\bl e$ are the vectorizations of
$\bl Y$, $\bl X$ and $\bl E$.
Both representations (\ref{eqn:blr}) and (\ref{eqn:vblr}) will be
useful in
what follows.
Note that the parameters $\bl A$ and $\bl B$ are not separately identifiable,
in that $\Exp{ \bl y | \bl x, c \bl A, \bl B /c} =
\Exp{ \bl y | \bl x, \bl A, \bl B }$ for any nonzero scalar $c$.
However, these parameters are identifiable up to scale, in the sense that
if $(\bl B\otimes\bl A ) \bl x = (\tilde{\bl B}\otimes\tilde{\bl A}
) \bl x$ for all $\bl x$, then $\tilde{\bl A} = c {\bl A} $ and
$\tilde{\bl B} = {\bl B}/c $ for some $c\neq0$ unless
all entries of either $\bl A$ or $\bl B$ are zero.

Given replications $\{ (\bl Y_1, \bl X_1),\ldots, (\bl Y_n, \bl X_n) \}$
from (\ref{eqn:blr}),
least squares parameter estimates $(\hat{\bl A},\hat{\bl B})$
of $(\bl A,\bl B)$
are minimizers of the residual mean squared error:
%
%
\begin{eqnarray}\label{eqn:lso}
(\hat{\bl A},\hat{\bl B} ) &=& \arg\min_{\bl A, \bl B} \sum
_{r=1}^n \bigl\Vert\bl Y_r - \bl A \bl
X_r \bl B^T \bigr\Vert^2/n
\nonumber
\\
&= &\arg\min_{\bl A, \bl B} \sum\Vert\bl Y_r
\Vert^2/n - 2 \sum\tr \bigl( \bl Y_r^T
\bl A \bl X_r \bl B^T/n \bigr)
\nonumber
\\[-8pt]
\\[-8pt]
\nonumber
&&{}+ \sum\tr \bigl(
\bl A \bl X_r \bl B^T \bl B \bl X_r \bl
A^T/n \bigr)
\\
&=& \arg\min_{\bl A, \bl B} \tr \Bigl( \bl A^T\bl A \sum
\bl X_r \bl B^T \bl B \bl X_r/n
\Bigr) -2 \tr \Bigl( \bl A^T \sum\bl Y_r \bl B
\bl X_r^T/n \Bigr), \nonumber
\end{eqnarray}
where $\tr(\bl H)$ denotes the trace of a square matrix $\bl H$, and the
term $\sum\Vert\bl Y_r\Vert^2/n$ has been dropped, as it does not
affect the minimization.
Equivalently, using representation (\ref{eqn:vblr}), we have
\begin{eqnarray*}
(\hat{\bl A},\hat{\bl B} ) &= &\arg\min_{\bl A, \bl B} \sum
_{r=1}^n\bigl \Vert\bl y_r - (\bl B \otimes
\bl A) \bl x_r \bigr\Vert^2/n
\\
&=& \arg\min_{\bl A, \bl B} \sum\Vert\bl y_r
\Vert^2/n - 2 \tr \Bigl( (\bl B\otimes\bl A) \sum\bl
x_r \bl y_r^T/n \Bigr) \\
&&{}+ \tr \Bigl( \bigl(
\bl B^T \bl B \otimes\bl A^T\bl A \bigr)\sum\bl
x_r\bl x_r^T/n \Bigr)
\\
&=& \arg\min_{\bl A, \bl B} f(\bl A,
\bl B, \bl S_{xx}, \bl S_{xy} ),
\end{eqnarray*}
where
%
%
\begin{equation}
f(\bl A, \bl B, \bl S_{xx}, \bl S_{xy}) = \tr \bigl( \bigl(
\bl B^T\bl B \otimes\bl A^T \bl A \bigr) \bl
S_{xx} \bigr) -2 \tr \bigl( (\bl B \otimes\bl A) \bl S_{xy}
\bigr), \label{eqn:vlso}
\end{equation}
with $\bl S_{xx} = \sum\bl x_r \bl x_r^T/n $ and
$\bl S_{xy} = \sum\bl x_r \bl y_r^T/n $.

Taking derivatives of the objective function
in (\ref{eqn:lso}) or (\ref{eqn:vlso})
with respect to $\bl A$ indicates that for a
nonzero value of $\bl B$, the minimizer of the residual mean squared
error in $\bl A$ is given
by
\[
\tilde{\bl A}(\bl B) = \Bigl( \sum\bl Y_r \bl B \bl
X_r^T \Bigr) \Bigl( \sum\bl X_r
\bl B^T \bl B \bl X_r^T \Bigr)^{-1}.
\]
A similar calculation shows that for a nonzero value of $\bl A$,
the minimizer in $\bl B$ is given by
\[
\tilde{\bl B}(\bl A) = \Bigl( \sum\bl Y_r^T
\bl A \bl X_r \Bigr) \Bigl( \sum\bl X_r^T
\bl A^T \bl A \bl X_r \Bigr)^{-1}.
\]
This suggests the following alternating least squares algorithm to
locate local minima of
(\ref{eqn:vlso}):
Given values
$\{ \hat{\bl A}^{(s)}, \hat{\bl B}^{(s)} \}$
at iteration $s$, new values are generated
as
$\hat{\bl A}^{(s+1)} = \tilde{\bl A} (\hat{\bl B}^{(s)} )$ and
$\hat{\bl B}^{(s+1)} = \tilde{\bl B} (\hat{\bl A}^{(s+1)} )$.
Such a procedure is a block coordinate descent algorithm, and
will converge to a local minimum of
(\ref{eqn:vlso}) if certain conditions on the data are met
[such as $\sum\bl X_r \bl B^T \bl B \bl X_r^T$ and
$\sum\bl X_r^T \bl A^T \bl A \bl X_r$
being invertible for all nonzero $\bl A$ and $\bl B$;
see \citet{luenbergerye2008}, Section~8.9].

One would hope that, given sufficient data, the parameter estimates would
bear some resemblance to the true data-generating mechanism.
We investigate this by examining the critical points of a large-sample
version of the objective function (\ref{eqn:vlso}).
Consider a scenario in which
$\bl S_{xx}= \sum\bl x_r \bl x_r^T/n $ converges almost surely
to a positive definite matrix $\Sigma_{xx} = \Exp{\bl x \bl x^T }$ and
$\bl S_{xy}= \sum\bl x_r \bl y_r^T/n $ converges almost surely to a matrix
$\Sigma_{xy}= \Exp{\bl x \bl y^T}$.
This implies almost sure convergence of
$f(\bl A, \bl B, \bl S_{xx}, \bl S_{xy})$
to $f(\bl A, \bl B, \Sigma_{xx}, \Sigma_{xy})$,
and so we would expect that
a minimizer of $f(\bl A, \bl B, \bl S_{xx}, \bl S_{xy})$
would resemble a minimizer of
$f(\bl A, \bl B, \Sigma_{xx}, \Sigma_{xy})$, given sufficient data.
In particular, results of \citet{white1981} imply that if
estimation
of $\{\bl A,\bl B\}$
is restricted to a compact subset
of $\mathbb R^{m_1\times p_1}\times\mathbb R^{m_2\times p_2}$,
then a sequence of local minimizers $\{ \hat{\bl A}_n, \hat{\bl
B}_n\}$
of $f(\bl A, \bl B, \bl S_{xx}, \bl S_{xy})$
will converge almost surely to
the global minimizer of
$f(\bl A, \bl B, \Sigma_{xx}, \Sigma_{xy})$, if one exists.
This motivates an investigation of minimizers
of $f(\bl A, \bl B, \Sigma_{xx}, \Sigma_{xy})$ under various
conditions on $\Sigma_{xx}$ and $\Sigma_{xy}$.
Such minimizers are referred to as ``pseudotrue'' parameters in
the literature on nonlinear least squares estimates and
misspecified models [see, e.g., \citeauthor{white1981}
(\citeyear{white1981,white1982})].

The ideal condition is, of course, when the model is correct.
In this case,
$\Exp{ \bl y | \bl x } = (\bl B_0 \otimes\bl A_0) \bl x$ and so
$\Sigma_{xy} = \Exp{\bl x \bl x^T (\bl B_0 \otimes
\bl A_0)^T } = \Sigma_{xx} (\bl B_0 \otimes\bl A_0)^T$.
The large-sample objective function
is then
\begin{eqnarray*}
&&f \bigl(\bl A, \bl B, \Sigma_{xx}, \Sigma_{xx} (\bl
B_0 \otimes\bl A_0)^T \bigr) \\
&&\qquad= \tr \bigl(
\bigl(\bl B^T\bl B \otimes\bl A^T\bl A \bigr)
\Sigma_{xx} \bigr) -2 \tr \bigl( (\bl B \otimes\bl A)
\Sigma_{xx} (\bl B_0 \otimes\bl A_0)^T
\bigr).
\end{eqnarray*}
If $\Sigma_{xx}$ is positive definite, then this
function is uniquely minimized
in $(\bl B \otimes\bl A)$ by the truth $(\bl B_0\otimes\bl A_0)$.
The pseudotrue parameters are equal to the true parameters, and
the least squares estimator
is asymptotically consistent.

If the model is incorrect, we may still hope that $(\hat{\bl A}, \hat
{\bl B})$ conveys meaningful information about the data-generating mechanism.
For example, recall that
$a_{i,j}$, the $i,j$th element of $\bl A$, represents a measure of the
conditional dependence of
$\bl y_{i} = (y_{i,1},\ldots, y_{i,m_2})^T$,
the $i$th row of $\bl Y$, on
$\bl x_j = ( x_{j,1},\ldots, x_{j,p_2})^T$,
the $j$th row of $\bl X$, given the other rows of $\bl X$.
If there is no such dependence,
then we would hope that the pseudotrue parameter for $a_{i,j}$ would be
zero as well.
It can be shown that this is true, under some additional conditions:

%
\begin{prop}\label{prop:ptrue1}
If
$\Exp{ \bl x_{j} \bl y_{i}^T } = \bl0$ and
$\Exp{ \bl x_{j} \bl x_{j'}^T } = \bl0$ for all $j'\neq j$, then
the pseudotrue
parameter for
$a_{i,j}$ is zero.
\end{prop}


A similar result holds
if the conditional expectation of $\bl Y$ given $\bl X$ is
truly linear, although not necessarily Kronecker structured. In this
case we can
write
$\Exp{\bl y | \bl x} = \bs\Theta\bl x$, where here $\bl y$ and $\bl
x$ are the vectorizations
of $\bl Y$ and $\bl X$.
%

\begin{prop}\label{prop:ptrue2}
Let $\Exp{ \bl y|\bl x} = \bs\Theta\bl x$ and
$\Exp{ \bl x \bl x^T } = \Omega\otimes\Psi$
for some positive definite matrices $\Omega$ and $\Psi$.
Then if the entries of $\bs\Theta$ corresponding to the elements of
$\bl y_{i}$ and $\bl x_{j}$
are zero, then the pseudotrue
parameter for
$a_{i,j}$ is zero.
\end{prop}

Proofs of both propositions are in Appendix \ref{appa}.
The conditions of both results
correspond to $\bl y_{i}$ being ``conditionally uncorrelated'' with
$\bl x_{j}$ in some way: Under the conditions of the first proposition,
the inverse of a covariance
matrix of the elements of $\bl Y$ and $\bl X$
would have zeros for all entries corresponding to
elements of $\bl y_{i}$ and $\bl x_{j}$, that is, the partial
correlations are
zero.
In the second proposition, $\bs\Theta$ represents the conditional
relationship directly.


\section{Extension to correlated multiway data}\label{sec3}
In this section the bilinear regression model is extended in two ways:
First, we show that the bilinear model is a special case of a more
general type of multilinear
tensor regression model that can be applied to tensor-valued data. Such
a model
can accommodate, for example, multivariate longitudinal relational data
of the
type described in Section~\ref{sec1}, where we have multiple relation
types measured between
pairs of countries over time. Such data can be represented as a time
series of
three-way tensors.
A second extension of the model allows for covariance in the error
term. As
sample size limitations will generally preclude
unrestricted estimation of the covariance, a~reduced-dimension
multilinear covariance model is proposed that
allows for correlation along each
mode of the tensor. The covariance model,
like the mean model, is obtained from a multilinear transformation,
so we refer to the combined mean and covariance model
as a general multilinear tensor regression
model (generalized MLTR).
The joint multilinear structure of the mean and covariance facilitates
parameter estimation.
In particular,
a~Bayesian approach to generalized least squares
(GLS) is available via a straightforward Gibbs sampling algorithm.

\subsection{Multilinear tensor regression}\label{sec3.1}
The bilinear regression model maps a
covariate matrix $\bl X\in\mathbb R^{p_1\times p_2}$
to a mean matrix $\bl M = \bl A \bl X \bl B^T \in\mathbb R^{m_1\times m_2}$.
Equivalently, the model
maps $\bl x$, the vectorization of $\bl X$,
to $\bl m =(\bl B \otimes\bl A ) \bl x$, the vectorization of $\bl M$.
Such a map between spaces of matrices is a special case of a
more general class of maps between spaces of multiway arrays, or
tensors. Specifically, given matrices $\bl B_1,\ldots, \bl B_K$, with
$\bl B_k \in\mathbb R^{m_k\times p_k}$, we can define a mapping
from $\mathbb R^{p_1\times\cdots\times p_K}$ to
$\mathbb R^{m_1\times\cdots\times m_K}$
by first obtaining the vectorization $\bl x$, computing
$\bl m = (\bl B_K \otimes\cdots\otimes\bl B_1) \bl x$, and then
forming an $m_1\times\cdots\times m_K$-dimensional array $\bl M$
from $\bl m$. This transformation is known as the ``Tucker product''
[\citet{tucker1964}] of the array $\bl X$ and the list of matrices $\bl
B_1,\ldots, \bl B_K$, which we write as
$\bl M = \bl X \times\{ \bl B_1,\ldots, \bl B_K\}$.

An important class of operations related to the Tucker product are \emph
{matricizations},
which reshape an array $\bl M$ into matrices of various dimensions.
For example, the mode-1 matricization of an $m_1\times m_2\times
m_3$-dimensional array $\bl M$ is an $m_1 \times(m_2
m_3)$-dimensional matrix
denoted $\bl M_{(1)}$.
More generally, the
mode-$k$ matricization of an $m_1\times\cdots\times m_K$-dimensional
array $\bl M$ is an $m_k \times( \prod_{k':k'\neq k}
m_{k'})$-dimensional matrix denoted $\bl M_{(k)}$.
The matricization operation
facilitates both understanding and computation of the
Tucker product via the following set of equivalencies:
%
%
\begin{eqnarray}
\bl M &=& \bl X \times\{ \bl B_1,\ldots, \bl B_K\},
\nonumber
\\
\bl m & =& (\bl B_K \otimes\cdots\otimes\bl B_1) \bl x,
\label{eqn:idvec}
\\
\bl M_{(k)} &=& \bl B_k \bl X_{(k)} ( \bl
B_{K} \otimes\cdots\otimes\bl B_{k+1} \otimes\bl
B_{k-1} \otimes\cdots\otimes\bl B_1 )^T.
\label{eqn:idmat}
\end{eqnarray}
In particular, (\ref{eqn:idmat}) can be used to compute the Tucker product
via a series of reshapings and matrix multiplications.
Additionally, this result indicates that the Tucker product consists of
a series of linear transformations along the different modes of the array.
More on the Tucker product and related operations can be found in, for example,
\citet{delathauwerdemoorvandewalle2000},
\citet{koldabader2009} and \citet{hoff2011b}.

Given an explanatory tensor $\bl X\in\mathbb R^{p_1\times\cdots
\times p_k}$
and an outcome tensor $\bl Y\in\mathbb R^{m_1\times\cdots\times m_K}$,
the Tucker product can be used to construct
a multilinear tensor regression model of the form
%
%
\begin{equation}
\bl Y = \bl X \times\{ \bl B_1,\ldots, \bl B_K\} + \bl
E, \label{eqn:mltr}
\end{equation}
where $\bl B_k \in\mathbb R^{m_k \times p_k}$, $k=1,\ldots,K$.
If $K=3$, for example, the model for element $i_1,i_2,i_3$ of
$\bl Y$ is
\[
y_{i_1,i_2,i_3} = \sum_{j_1} \sum
_{j_2} \sum_{j_3} b_{1,i_1,j_1}
b_{2,i_2,j_2} b_{3,i_3,j_3} x_{j_1,j_2,j_3} + \varepsilon_{i_1,i_2,i_3},
\]
and so $b_{1,i_1,j_1}$ can be viewed as the multiplicative effect
of ``slice'' $j_1$ of $\bl X$ on slice $i_1$ of $\bl Y$.
The similarity of this model to the bilinear regression model
is most easily seen via the vectorized version of
(\ref{eqn:mltr}), which takes the following form:
%
%
\begin{equation}
\bl y = (\bl B_K \otimes\cdots\otimes\bl B_1 ) \bl x +
\bl e. \label{eqn:vmltr}
\end{equation}

With this notation, replicate observations
$\{ (\bl Y_1,\bl X_1),\ldots, (\bl Y_n,\bl X_n)\}$
are easily handled by ``stacking'' the arrays to
form two $(K+1)$-way arrays $\bl Y\in\mathbb R^{m_1\times\cdots
\times m_K \times n}$ and
$\bl X\in\mathbb R^{p_1\times
\cdots\times p_K \times n}$, where the $(K+1)$st mode indexes the
replications. If each slice follows
model (\ref{eqn:mltr}), then the model for the stacked data is
%
%
\begin{eqnarray}\label{mltrrep}
\bl Y &=& \bl X \times\{ \bl B_1,\ldots, \bl B_K, \bl
I_n\} + \bl E \ \quad\mbox{or, equivalently,}
\nonumber
\\[-8pt]
\\[-8pt]
\nonumber
\bl y &=&(\bl I_n \otimes\bl B_k \otimes\cdots\otimes\bl
B_1 ) \bl x + \bl e,
\end{eqnarray}
where $\bl I_n$ is an $n\times n$ diagonal matrix,
$\bl E$ is a mean-zero array of the same dimension as $\bl Y$,
and $\bl e$ is the vectorization of $\bl E$.
However,
in what follows we work with model~(\ref{eqn:mltr}), while recognizing that
estimation with replications can be handled as a special case by
stacking the replications and fixing the parameter
matrix for the last mode to be the identity matrix.

Estimation is facilitated by application of
identity (\ref{eqn:idmat}).
For example, matricizing
each term in (\ref{eqn:mltr}) along the first mode gives
%
%
\begin{eqnarray}
{\bl Y}_{(1)} & = \bl B_1 \tilde{\bl X}_{(1)} + {
\bl E}_{(1)}, \label{eqn:redform}
\end{eqnarray}
%
where
$\tilde{\bl X}_{(1)} =\bl X_{(1)} (\bl B_K \otimes\cdots\otimes\bl B_2)^T$.
In terms of $\bl B_1$, this is simply a multivariate linear regression model
[\citet{mardiakentbibby1979}, Chapter~6].
The least squares criterion in $\bl B_1$ is $\Vert{\bl Y} - \bl B_1
\tilde{\bl X}\Vert^2$, which is uniquely minimized in
$\bl B_1$ by ${\bl Y} \tilde{\bl X}^T ( \tilde{\bl X} \tilde{\bl
X}^T)^{-1}$ (if $\tilde{\bl X}$ has full row rank).
Similar forms result from
matricizing along any of the other $K$ modes.
It follows that estimates of
$\{\bl B_1,\ldots, \bl B_K\}$ can be obtained by generalizing the
block coordinate descent algorithm described in Section~\ref{sec2}.
Given starting values of $\bl B_1,\ldots, \bl B_K$, the algorithm is to
iterate the following steps
until convergence:

For for each $k\in\{1,\ldots, K\}$:
\begin{longlist}[1.]
\item [1.]compute $\tilde{\bl X} = \bl X \times\{ \bl B_1,\ldots,\bl
B_{k-1}, \bl I_{p_k},\bl B_{k+1},\ldots, \bl B_K \}$;
\item[2.] form ${\bl Y}_{(k)}$ and $\tilde{\bl X}_{(k)}$, the mode-$k$
matricizations of $\bl Y$ and $\tilde{\bl X}$;
\item[3.] set $\bl B_k = {\bl Y}_{(k)} \tilde{\bl X}_{(k)}^T
( \tilde{\bl X}_{(k)} \tilde{\bl X}_{(k)}^T )^{-1}$.
\end{longlist}
Note that in the algorithm we are computing $\tilde{\bl X}_{(1)}$,
for example,
by first computing $\bl X \times\{ \bl I_{p_1}, \bl B_2, \ldots, \bl
B_K \}$ and then matricizing,
rather than matricizing $\bl X$ and then multiplying on the right by
$(\bl B_K \otimes\cdots\otimes\bl B_2)^T$.
The two approaches give the same result, but the former can be accomplished
with $K-1$ ``small'' matrix multiplications, whereas the
latter requires construction of and multiplication by
$(\bl B_K \otimes\cdots\otimes\bl B_2)^T$, which can be unmanageably large
in some applications.

%
\begin{figure}

\includegraphics{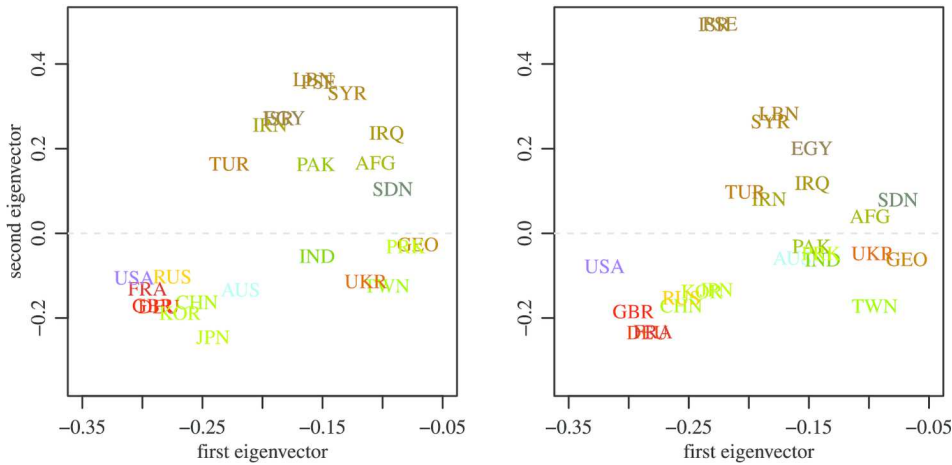}

\caption{Eigenvectors of mode-specific residual correlation matrices.}
\label{fig:rescor}
\end{figure}

\subsection{Inference under a separable covariance model}\label{sec3.2}

The international relations data
presented in Section~\ref{sec1}, and
that will be more fully analyzed in Section~\ref{sec4},
consist of time series of four different relational measurements
between pairs of 25 countries. These data can be represented as
a four-way array $\bl Y \in\mathbb R^{25\times25\times4 \times543}$.
Using the algorithm described in Section~\ref{sec3.1}, least squares
estimates of $\{ \bl B_1,\bl B_2,\bl B_3\}$ for the
model $\bl Y = \bl X \times\{ \bl B_1,\bl B_2, \bl B_3, \bl I \}
+\bl E$
were obtained,
where $\bl X$ is a lagged version of $\bl Y$.
These estimates are equivalent to
maximum likelihood estimates
under the assumption of
i.i.d. residual variation.
The plausibility of this assumption is examined graphically in
Figure~\ref{fig:rescor}. This plot
shows eigenvectors of the sample correlation matrices of
$\bl R_{(1)} $ and $\bl R_{(2)}$, which are the mode-1 and mode-2
matricizations of the residual array
$\bl R = \bl Y - \bl X \times\{ \hat{\bl B}_1,\hat{\bl B}_2, \hat{\bl
B}_3, \bl I \} $.
These plots should appear patternless under the assumption of
i.i.d. residuals. Instead, clear patterns
of residual correlation among certain groups of countries
are exhibited, many of which
are geographic.
In cases like this, where
residual variation is not
well represented by an i.i.d. model, it may be preferable
to use an estimation method that accounts for residual
correlation or heteroscedasticity.


Given multiple observations,
we might model the residuals in the vectorized version of the model
(\ref{eqn:vmltr}) as
$\bl e_1,\ldots, \bl e_n \sim\mbox{i.i.d. } N_{m}(\bl0, \Sigma)$,
where
$m=\prod m_k$ and
$\Sigma$ is an unknown covariance matrix to be estimated.
The difficulty with this, as with an unrestricted regression model, is that
the sample size will generally be too small to reliably estimate
$\Sigma$ without making some restrictions on its form.
A flexible, reduced-parameter covariance model that retains
the tensor structure of the data is the array normal model
[\citet{akdemirgupta2011},
Hoff (\citeyear{hoff2011b})], which assumes a
separable (Kronecker structured) covariance matrix.
For example, we say that $\bl E$ has a mean-zero array normal distribution,
and write
$\bl E \sim N_{m_1\times\cdots\times m_K}(\bl0, \Sigma_1,\ldots,
\Sigma_K)$, if the distribution of the vectorization $\bl e$ of $\bl E$
is given by
$ \bl e \sim N_{m}(\bl0,
\Sigma_K \otimes\cdots\otimes\Sigma_1), $
where
$\Sigma_k$ is a positive definite $m_k\times m_k$ matrix for
each $k=1,\ldots, K$.
Each $\Sigma_k$ can be interpreted as the covariance along the $k$th
mode of $\bl E$.
For example,
if $\bl E \sim N_{m_1\times\cdots\times m_K}(\bl0,
\Sigma_1, \ldots, \Sigma_K)$,
then it is straightforward to
show that
$\Exp{ \bl E_{(k)} \bl E_{(k)}^T } \propto\Sigma_k$,
where $\bl E_{(k)}$ is the mode-$k$ matricization of~$\bl E$.

Combining this error model with the mean model in (\ref{eqn:mltr}),
and applying identities (\ref{eqn:idvec}) and (\ref{eqn:idmat}),
gives three equivalent forms for this general multilinear
tensor regression model:
%
%
\begin{eqnarray}\label{eqn:mlmforms}
&& \mbox{Tensor form:}\quad \bl Y = \bl X \times\{ \bl
B_1,\ldots, \bl B_K\} + \bl E,\nonumber\\
  \eqntext{\bl E \sim
N_{m_1\times\cdots\times m_K }( \bl0, \Sigma_1,\ldots, \Sigma_K),}
\\
&&\mbox{Vector form:}\quad  \bl y = ( \bl B_K \otimes\cdots\otimes\bl
B_1) \bl x + \bl e,
\nonumber
\\[-8pt]
\\[-8pt]
  \eqntext{\bl e\sim N_{m}(\bl0,
\Sigma_K \otimes\cdots\otimes\Sigma_1 ),}
\\
&&\mbox{Matrix form:}\quad  \bl Y_{(k)} = \bl B_k \bl
X_{(k)} \bl B_{-k}^T + \bl E_{(k)},\nonumber\\
  \eqntext{\bl
E_{(k)} \sim N_{m_k \times m_{-k}}(\bl0, \Sigma_k,
\Sigma_{-k} ),}
\end{eqnarray}
where in the matrix form,
$\bl B_{-k}=\bl B_K \otimes\cdots\otimes\bl B_{k+1} \otimes\bl
B_{k-1} \otimes\cdots\otimes\bl B_1$,
$\Sigma_{-k}$ is defined similarly
and $m_{-k} = \prod_{k':k'\neq k} m_{k'}$.
As before, we note that $n$ replications from a $K$-mode model can be
represented by stacking the data arrays and using
a $(K+1)$-mode model with the restriction that $\bl B_{K+1} = \Sigma
_{K+1} = \bl I_n$.

As in the uncorrelated case,
the matrix form of the model can be used to obtain iterative algorithms
for parameter estimation. For example,
multiplying the terms in the matrix form on the right by
$\Sigma_{-k}^{-1/2}$ allows us to express the model as
%
%
\begin{equation}
\tilde{\bl Y}_{(k)}  = \bl B_k \tilde{\bl
X}_{(k)} + \tilde{\bl E}_{(k)},\qquad \tilde{\bl E}_{(k)}
\sim N_{m_k\times m_{-k} } (\bl0, \Sigma_k,\bl I_{m_{-k}}),
\label{eqn:credform}
\end{equation}
where now $\tilde{\bl Y}_{(k)} = \bl Y_{(k)} \Sigma_{-k}^{-1/2}$ and
$\tilde{\bl X}_{(k)} =\bl X_{(k)} \bl B_{-k}^T \Sigma_{-k}^{-1/2}$.
Given the parameters other than $\bl B_k$, this is a multivariate
linear regression model
with dependent errors.
The (conditional) MLE and generalized least
squares estimator is
$\hat{\bl B}_k = \tilde{\bl Y}_{(k)} \tilde{\bl X}_{(k)}^T
(\tilde{\bl X}_{(k)}\tilde{\bl X}_{(k)}^T )^{-1}$, which has the same form
as the OLS estimator [see, e.g., \citet{mardiakentbibby1979},
Section~6.6.3], except here the covariance along the modes other than $k$
have been incorporated into the construction of $\tilde{\bl Y}_{(k)}$ and
$\tilde{\bl X}_{(k)}$.
Generalized least squares estimates of the $\bl B_k$'s, conditional on
values of the
$\Sigma_k$'s, can thus be found via the coordinate descent algorithm in
the previous subsection, modulo the modification to $\tilde{\bl
Y}_{(k)}$ and $\tilde{\bl X}_{(k)}$. Analogously, given current values
of the $\bl B_k$'s,
the likelihood can be minimized in the $\Sigma_k$'s by applying
a similar iterative algorithm, described in
\citet{hoff2011b}.

\subsection{Bayesian estimation and inference}\label{sec3.3}
Generally speaking, maximum likelihood estimates
in high-dimensional settings can be unstable and overfit to the data.
Such problems can often be ameliorated by
instead obtaining estimates that maximize a
penalized likelihood.
By viewing a penalty as a prior distribution,
penalized estimates can be obtained via
Bayesian procedures,
which have the additional advantage of providing a very complete
description of parameter uncertainty.
In particular, Markov chain Monte Carlo (MCMC) methods that
approximate posterior distributions are useful for exploring the
parameter space in a way that is often more informative
than computing a matrix of second derivatives at a local mode,
especially if the dimension of the parameter space is large.
With this in mind, we present a class of semiconjugate prior
distributions for the model (\ref{eqn:mlmforms}), and obtain a Gibbs
sampler that can be used to simulate parameter values from the
corresponding posterior distribution.

Recall from the previous subsection that, given $\{\bl B_{k'}:k'\neq k\}$
and $\{ \Sigma_{k'}: k'\neq k\}$, the model in terms of $(\bl B_k,
\Sigma_k)$
can be expressed as an ordinary multivariate regression model,
%
%
\begin{equation}
\bl Y \sim N_{m\times n} (\bl B \bl X, \Sigma, \bl
I_n), \label{eqn:mlreg}
\end{equation}
where $\bl B\in\mathbb R^{m\times p}$ and
$\Sigma\in\mathcal S_p^+$ are to be estimated
from
$\bl Y\in\mathbb R^{m\times n}$ and $\bl X\in\mathbb R^{p\times n}$.
As such,
Bayesian inference for $\{(\bl B_k, \Sigma_k), k=1,\ldots, K\}$
can be made via a Gibbs sampler that iteratively re-expresses
the model in terms of (\ref{eqn:mlreg}) for each mode $k$, and simulates
$(\bl B_k, \Sigma_k)$ from the corresponding posterior distribution.

Posterior inference for (\ref{eqn:mlreg}) is facilitated by choosing
a conjugate prior, which for this model is
$\Sigma\sim\operatorname{inverse\mbox{-}Wishart}( \bl S_0^{-1}, \nu_0)$ and
$\bl B|\Sigma\sim N_{m\times p}( \bl M_0,\break  \Sigma, \bl I_p )$,
where the inverse-Wishart distribution is parameterized so that\break 
$\Exp{ \Sigma^{-1} } = \nu_0 \bl S_0^{-1}$.
Under this prior and model (\ref{eqn:mlreg}), the joint
posterior density of $(\bl B, \Sigma)$ given $\bl Y$ can be expressed
as $p(\bl B, \Sigma|\bl Y) = p(\bl B | \Sigma, \bl Y) \times p(\bl
\Sigma|\bl Y) $, where the first density on the right-hand side is a
matrix normal
density, and the second is an inverse-Wishart density. Specifically,
%
%
\begin{eqnarray}\label{eqn:sigpost}
\Sigma|\bl Y &\sim&\operatorname{inverse\mbox{-}Wishart} \bigl(\bl S_n^{-1},
\nu_0 + n \bigr)
\nonumber
\\[-8pt]
\\[-8pt]
\eqntext{\mbox{where } \bl S_n=\bl
S_0 +\bl Y \bigl(\bl I_n+\bl X^T \bl X
\bigr)^{-1} \bl Y^T ;}
\\
 \label{eqn:bpost}
 \bl B| \Sigma, \bl Y& \sim &N_{m\times p} \bigl(\bl M_n, \Sigma,
\bigl(\bl I_p + \bl X \bl X^T \bigr)^{-1}
\bigr)
\nonumber
\\[-8pt]
\\[-8pt]
\eqntext{\mbox{where } \bl M_n = \bigl( \bl M_0 + \bl Y
\bl X^T \bigr) \bigl( \bl I_p + \bl X \bl X^T
\bigr)^{-1}.}
\end{eqnarray}
%
Typically, $n$ will be much larger than $p$, in which case
$\bl S_n$ is more efficiently calculated as
$\bl S_n = \bl S_0 + \bl Y ( \bl I_n - \bl X^T (\bl I+\bl X \bl
X^T)^{-1} \bl X) \bl Y^T$, which requires inversion of a $p\times p$
matrix rather than an $n\times n$ matrix.

Returning to the tensor regression model, for Bayesian analysis
we parameterize the model as
\begin{eqnarray*}
\bl Y & = &\bl X \times\{ \bl B_1,\ldots, \bl B_K \} +
\tau\bl E,
\\
\bl E &\sim & N_{m_1\times\cdots\times m_K } ( \bl0, \Sigma_1,\ldots,
\Sigma_K ),
\end{eqnarray*}
where $\tau$ is an additional scale parameter that decouples the
magnitude of the error variance from the prior variance of the $\bl B_k$'s
(both of which would otherwise be determined by
the $\Sigma_k$'s).
An $\operatorname{inverse\mbox{-}gamma}(\eta_0/2,\eta_0 \tau_0^2/2)$ prior distribution for
$\tau^2$ results in an
$\operatorname{inverse\mbox{-}gamma}([\eta_0+m]/2, [ \eta_0\tau_0^2+ \Vert\bl Y - \bl
X\times\{ \Sigma_1^{-1/2}\bl B_1,\ldots,\Sigma_K^{-1/2} \bl B_K\}
\Vert^2 ] /2 )$
full conditional distribution.
Based on these results, a~Gibbs sampler with a stationary distribution
equal to the posterior distribution of $\{ \bl B_1,\ldots,\bl B_K$,
$\Sigma_1,\ldots,\Sigma_K$, $\tau^2\}$ can be constructed by iterating
the following steps:
\begin{enumerate}[1.]
\item[1.] Iteratively for each $k=1,\ldots, K$:
\begin{enumerate}[(a)]
\item[(a)] compute $\tilde{\bl Y} = \bl Y_{(k)} \bl\Sigma^{-1/2}_{-k}
/\tau$
and $\tilde{ \bl X} = \bl X_{(k)} \bl B_{-k}^T \bl\Sigma
^{-1/2}_{-k}/\tau$;
\item[(b)] simulate $(\Sigma_k, \bl B_k)$ from
(\ref{eqn:sigpost}) and (\ref{eqn:bpost}),
replacing $\bl Y$
and $\bl X$ with $\tilde{\bl Y}$ and $\tilde{\bl X}$.
\end{enumerate}
\item[2.] Simulate $\tau^2\sim
\operatorname{inverse\mbox{-}gamma}( [\eta_0+m]/2, [ \eta_0\tau_0^2+ \Vert\bl Y -
\bl X\times\{ \Sigma_1^{-1/2}\bl B_1,\ldots,\break  \Sigma_K^{-1/2} \bl B_K\}
\Vert^2 ] /2 )$.
\end{enumerate}
Parameter values simulated from this Markov chain can be used to make
Monte Carlo approximations to posterior quantities of interest.

\section{Analysis of longitudinal multirelational IR data}\label{sec4}
In this section we analyze weekly counts of four different
action types between 25 countries
over the ten and a half-year period from 2004 through the middle of 2014.
These data were obtained from the
ICEWS project
(\url{http://www.lockheedmartin.com/us/products/W-ICEWS/iData.html}),
which records
time-stamped
actions taken by one country
with another country as the target.
The 25 countries included in this analysis consist of the most
active countries during the time period.
The action types correspond to the four
``quad classes'' often used in international relations
event analysis, and include
negative material actions,
positive material actions, negative
verbal actions and positive verbal actions,
denoted m$-$, m$+$, v$-$, v$+$, respectively.
Examples of events that would fall into each of these four categories
are as follows:
imposing a blockade (m$-$),
providing humanitarian aid (m$+$),
demanding a change in leadership (v$-$),
and granting diplomatic recognition (v$+$).
These data can be expressed as a $25\times25 \times4 \times
543$-dimensional array $\bl Y$, where entry $y_{i_1,i_2,j,t}$
corresponds to the number of
actions of type $j$, taken by country $i_1$ with country $i_2$ as the target,
during week $t$.
A normal quantile--quantile transformation was applied to each time series
corresponding to an actor-target-type triple, so that
for each $i_1,i_2,j$, the empirical distribution of
$\{ y_{i_1,i_2,j,t}: t=1,\ldots, 543\}$
is approximately standard normal.

This section presents several candidate models for these data,
and presents in detail the estimation results for the one providing
the best fit in terms of predictive $R^2$.
Perhaps the simplest modeling approach is to fit four separate
bilinear regression models to each of the four action types, that is,
to fit
$\bl Y^{(j)} = \bl X^{(j)}\times\{ \bl B_1^{(j)}, \bl B_2^{(j)}, \bl
I \} + \bl E^{(j)}$, where $\bl Y^{(j)}$ is the
$25\times25 \times543$ array of between-country relations of type $j$,
and $\bl X^{(j)}$ is a lagged version of $\bl Y^{(j)}$,
for each $j\in\{1,\ldots, 4\}$.
A competing model is the joint multilinear model
$\bl Y = \bl X \times\{ \bl B_1,\bl B_2, \bl B_3,\bl I\}+\bl E$,
where $\bl Y$ is the complete $25\times25 \times4 \times543$ data
array, and $\bl B_3$ is a $4\times4$ matrix of coefficients
representing the effects of the different event types on one another.
One possible advantage of using separate bilinear fits is that
separate coefficient matrices $\bl B_1$ and $\bl B_2$ can be estimated for
each event type. Two disadvantages of this approach, as compared to the
joint multilinear procedure, are that (1) the bilinear approach
does not make use of one relation type to help predict another, and
(2) if the coefficient matrices are
not substantially different across event types, then
fitting them to be equal
(as in the
multilinear model) could improve estimation.

%
\begin{table}
\tabcolsep=0pt
\caption{Averages (and ranges in parentheses) of predictive
$R^2$-values across the ten cross-validation data sets,
for each model}
\label{tab:cvstudy}
\begin{tabular*}{\textwidth}{@{\extracolsep{\fill}}lcccc@{}}
\hline
\textbf{Model} & \textbf{Material}$\bolds{-}$ & \textbf{Material}$\bolds{+}$ &
\textbf{Verbal}$\bolds{-}$ & \textbf{Verbal}$\bolds{+}$ \\
\hline
Separate bilinear &7.9 (7.0, 9.5) & 2.9 (1.8, 3.5) &
7.8 (6.9, 9.0)\phantom{0} &
12.3 (10.9, 13.7) \\
Joint multilinear & \phantom{0}8.9 (7.9, 10.3) & 3.6 (2.9, 4.4) &
9.5 (8.5, 11.1) &
12.5 (11.5, 13.7) \\
Relational multilinear & 11.0 (9.6, 12.6) & 4.5 (3.5, 5.0) & 11.5
(10.7, 12.9) & 13.6 (12.6, 14.7)\\
\hline
\end{tabular*}
\end{table}

Inspection of the OLS estimates of
$\{ (\bl B_1^{(j)}, \bl B_2^{(j)}), j=1,\ldots, 4\}$
indicated a high degree of similarity across the four action
types, suggesting that the joint multilinear model may be appropriate.
More formally, we compared the separate and joint models using a
10-fold cross-validation study as described in the \hyperref
[sec1]{Introduction}:
For each of the 10 training sets,
OLS estimates for each model were obtained using the
algorithm described in Section~\ref{sec3.1}.
Averages of predictive $R^2$-values, as well as their ranges across the
10 test sets, are presented in
Table~\ref{tab:cvstudy}.
The results indicate that, in terms of out-of-sample predictive performance
for each action type,
the benefits of the joint multilinear model outweigh the flexibility of
having separate bilinear fits.

\subsection{Reciprocity and transitivity}\label{sec4.1}
We now extend the explanatory tensor $\bl X$ to account for certain
types of patterns
often seen in relational data and social networks.
One such pattern is the tendency for actions from
one node $i_1$ to another node $i_2$ to be reciprocated over time,
so that if $y_{i_1,i_2,j,t}$ is large,
we may expect $y_{i_2,i_1,j,t+1}$ to be large as well.
To estimate such an effect from the data,
we add four ``slices'' to the tensor $\bl X$
along its third mode as follows:
Redefine $\bl X$ so that
$\bl X \in\mathbb R^{25\times25 \times8 \times543}$,
with lagged elements $x_{i_1,i_2,j,t} = y_{i_1,i_2,j,t-1}$ for $j\in\{
1,\ldots, 4\}$
as before,
and
reciprocal lagged elements
$x_{i_1,i_2,j,t} = y_{i_2,i_1,j-4,t-1}$ for $j\in\{5,\ldots, 8\}$.
A multilinear regression model of the form
$\bl Y = \bl X \times\{ \bl B_1, \bl B_2, \bl B_3, \bl I \} + \bl E$
then has $\bl B_3 \in\mathbb R^{4\times8}$, the
first four columns of which describe, for example, the effects of
$y_{i_1,i_2,j,t-1}$ on $y_{i_1,i_2,j,t}$, and the last four columns
of which describe the effects of
$y_{i_2,i_1,j,t-1}$ on $y_{i_1,i_2,j,t}$, that is, the tendencies of
actions to be reciprocated 
at the
next time point.

Other network effects can be accommodated similarly. One common
pattern in network and relational data is a type of third-order
dependence known as transitivity, which describes how
the simultaneous presence of relations between nodes $i_1$ and $i_3$, and
between $i_2$ and $i_3$, might lead to a relation from $i_1$ to $i_2$.
Based on this idea,
we construct a transitivity
predictor for each action type and add them to the third mode
of $\bl X$. Specifically,
we let $x_{i_1,i_2,j,t} = \sum_{i_3} ( y_{i_1,i_3,j-8,t} + y_{i_3,i_1,j-8,t})
( y_{i_2,i_3,j-8,t} + y_{i_3,i_2,j-8,t})$ for each
$j\in\{9,10,11,12\}$,
so that now $\bl X \in\mathbb R^{25\times25\times12\times543}$,
and the last four columns of the coefficient matrix $\bl B_3\in\mathbb
R^{4\times12}$
represent how the relations of nodes $i_1$ and $i_2$ with
common targets
lead to actions between $i_1$ and $i_2$ at the next time point.
Note that this is a simplified measure of transitivity, in that
the directions of the actions are not accounted for.
In what follows,
we refer to this regression model as a relational multilinear regression,
as it includes terms that allow estimation of patterns of reciprocity
and transitivity that are often observed in relational data.

\subsection{Longer-term dependence}\label{sec4.2}
Finally, we illustrate how to extend the relational multilinear model
to account for longer-term
longitudinal dependence. The appropriateness of doing so
for these data
is suggested by
Figure~\ref{fig:exdata}: While the week-$t$
observations are predictive of those at week $t+1$,
some trends in the time series appear to persist beyond one week.
In a separate exploratory analysis
(not presented here), we considered
using lagged monthly averages as predictors, along with the one-week lag
currently in the model.
We found that after including a one-week lag and
a one-month lag (the latter being an average
of four weeks of previous data),
the effects of lagged data from earlier months were minimal. For this
reason, in what follows we model the data at time $t+1$
as a function of
the data from the previous week $t$, as well as the average of
the data from the previous month (weeks $t-1,t-2,t-3,t-4$).

One possibility for incorporating the one-month lagged data
would be to add 12 more variables along the third mode of $\bl X$ as in
the previous subsection. Each of these 12 variables would represent a
monthly lagged version
of the existing 12 variables along this mode. Such an approach
would double the dimension of $\bl B_3$ and also make the interpretation
of parameter values more cumbersome. A more parsimonious alternative is
to assume separability of the effects of the two lag scales (weekly and
monthly). Specifically, we reconstruct $\bl X$ to be a
$25\times25 \times12 \times2 \times543$-dimensional tensor,
where $x_{i_1,i_2,j,1,t}$ corresponds to the previously existing
entries of
$\bl X$, and
$x_{i_1,i_2,j,2,t}$ corresponds to the
average of $x_{i_1,i_2,j,1,t-1},\ldots,
x_{i_1,i_2,j,1,t-4}$, that is, the average of the
previous month's predictors.
Treating $\bl Y$ as a $25\times25\times4 \times1 \times543$
dimensional array, the multilinear regression model of $\bl Y$ on $\bl X$
is expressed as
%
%
\begin{equation}
\bl Y = \bl X \times\{ \bl B_1,\bl B_2, \bl
B_3, \bl B_4, \bl I\} +\bl E, \label{eq:fmod}
\end{equation}
where $\bl B_4$ is a $1\times2$ matrix (or vector) that describes
the effect of 1-week lagged data relative to that of the 1-month
lagged data.

\subsection{Parameter estimation and interpretation}\label{sec4.3}
We first compare the predictive performance of
the least squares estimates from the relational multilinear model~(\ref{eq:fmod}) to the performance of the previously discussed models,
using the 10-fold cross-validation procedure described above.
As shown in Table~\ref{tab:cvstudy}, model (\ref{eq:fmod})
outperforms the
others in terms of predictive performance, and
in fact outperformed 
the joint multilinear model on each of the 10
test data sets.
These results suggest that this model is not overfitting relative to these
simpler
models.


%
\begin{figure}

\includegraphics{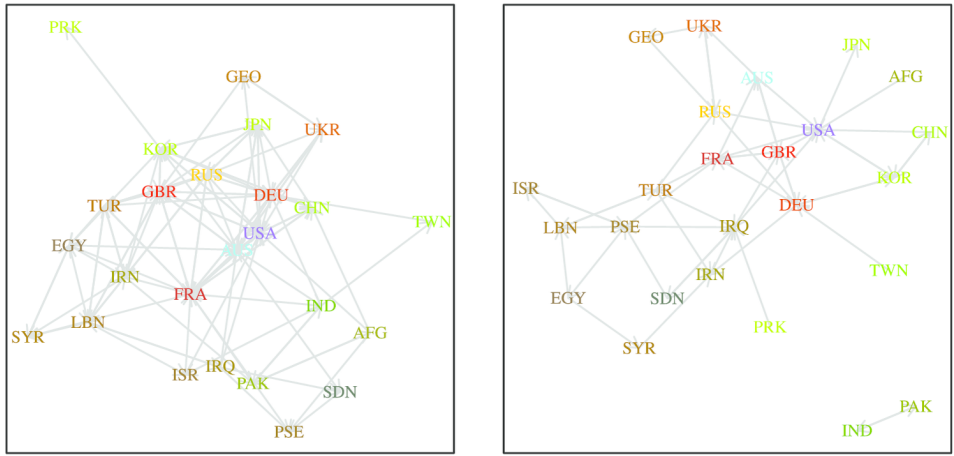}

\caption{Summary of the posterior distributions of $\bl B_1$ and $\bl
B_2$.}
\label{fig:infnet}
\end{figure}

A more complete description of these data can be obtained
via a Bayesian analysis
of (\ref{eq:fmod}), using a separable model
for residual covariance
as described in Section~\ref{sec3.2}.
Such an analysis accommodates
residual dependence and provides an assessment of parameter uncertainty
using, for example, Bayesian confidence intervals.
For this analysis,
we used diffuse but proper priors,
with $(\nu_0,\tau_0^2)=(1,1)$,
and for each mode $k$, $\bl M_{0k} = \bl0$,
$\bl S_{0k} = \bl I_{m_k}$ and
$\nu_{0k}= m_k + 1$.
The resulting posterior distributions of $\bl B_1$ and $\bl B_2$ are
summarized in
Figure~\ref{fig:infnet}.
(Details on the MCMC approximation are provided in the \hyperref
[app]{Appendix}.)
In each panel,
nominally significant positive effects are shown by drawing
a directed link from country $i_1$ to country $i_2$ if
the lower 99\% posterior quantile for entry
$i_1,i_2$ of $\bl B_1$ or $\bl B_2$ is greater than zero
(the 99th quantile
was used instead of the 95th to ensure readability of the graphs).
Also, there were
very few negative coefficients of $\bl B_1$ and $\bl B_2$:
only approximately 1\% had their upper 99\% posterior quantile below zero.
Not shown in the graph is that the lower 99\% posterior quantile
of each diagonal entry of $\bl B_1$ and $\bl B_2$ was positive,
and that these coefficients were generally much larger in magnitude than
the off-diagonal coefficients:
For example, the diagonal elements of the posterior mean of $\bl B_1$
were about 35 times larger than its off-diagonal elements, on average.
The diagonal elements of $\hat{\bl B}_3$ were also larger than
the off-diagonal elements (as shown in Table~\ref{tab:b3}), but to a lesser
extent.
These results indicate that, in general, the strongest predictor of
$y_{i_1,i_2,j,t}$ is $x_{i_1,i_2,j,t}$.
The next strongest predictors generally include
$x_{i_1,i_2,j',t}$ (a relation of a different type between the same dyad),
then
$x_{i',i_2,j,t}$ or $x_{i_1,i',j,t}$
(relations involving either the same actor or the same target)
depending on whether or
not $\hat b_{1,i_1,i'}$ or $\hat b_{2,i_2,i'}$ is moderately large.
Interpretation may be further aided with the
following example: Letting $i_1$ denote the
index of Iran, for example, the largest value of
$\{ b_{1,i_1,i'}: i'\in\{1,\ldots, 25\}\setminus\{i_1\} \}$
corresponds to that of Syria. The parameter estimates thus predict that
actions of Syria toward a country $i_2$ will increase the probability
of actions of Iran toward $i_2$, at a future time point.
Posterior means and standard deviations for the top ten nondiagonal
elements of $\bl B_1$ and $\bl B_2$, in terms of the ratio of
mean to standard deviation, are given in
Table~\ref{tab:B12_esd}.

%
\begin{table}
\caption{Posterior means and standard deviations of the top ten elements of
$\bl B_1$ and $\bl B_2$, in terms of
the ratio of mean to standard deviation}
\label{tab:B12_esd}
\begin{tabular*}{\textwidth}{@{\extracolsep{\fill}}lcclcc@{}}
\hline
\multicolumn{3}{c}{$\bolds{\bl B_1}$} & \multicolumn{3}{c@{}}{$\bolds{\bl B_2}$} \\[-6pt]
\multicolumn{3}{@{}c}{\hrulefill} & \multicolumn{3}{c@{}}{\hrulefill} \\
$\bolds{i_1,i_2}$ & $\mathbf{E}\bolds{[b_{1,i_1,i_2}]}$ &
$\mathbf{SD}\bolds{[ b_{1,i_1,i_2}]}$ &
$\bolds{i_1,i_2}$ & $\mathbf{E}\bolds{[b_{2,i_1,i_2}]}$ &
$\mathbf{SD}\bolds{[ b_{2,i_1,i_2}]}$ \\
\hline
GBR DEU &0.137& 0.023& GBR DEU& 0.110& 0.022 \\
DEU FRA &0.121& 0.018& GBR AUS& 0.101& 0.024 \\
TUR IRN &0.120& 0.015& ISR PSE& 0.092& 0.022 \\
FRA DEU &0.120& 0.021& IRQ USA& 0.067& 0.012 \\
JPN KOR &0.114& 0.020& AUS GBR& 0.066& 0.014 \\
AUS GBR &0.097& 0.016& RUS USA& 0.063& 0.013 \\
GBR USA &0.096& 0.012& GBR USA& 0.060& 0.012 \\
LBN IRN &0.088& 0.012& LBN ISR& 0.060& 0.014 \\
KOR CHN &0.088& 0.015& PRK IRQ& 0.054& 0.011 \\
UKR RUS &0.061& 0.011& SDN IRQ& 0.047& 0.011 \\
\hline
\end{tabular*}
\end{table}

The posterior distribution of the $\bl B_3$
coefficients, which describe the main, reciprocal and transitive effects
of the four action types on future actions,
is summarized in Table~\ref{tab:b3}.
This table gives posterior mean estimates of those
coefficients of $\bl B_3$ for which zero is not included in their 95\%
posterior confidence interval.
The first four columns of this matrix largely represent
the direct effects of
action variable $j_1$ from $i_1$ to $i_2$ on the future value of
action variable $j_2$ from $i_1$ to $i_2$, for $j_1,j_2 \in\{1,2,3,4\}$.
Not surprisingly, the largest estimated coefficients are along the diagonal,
indicating that the strongest predictor of action variable $j_1$ is
the previous value of
this variable. Other ``significant'' coefficients include
effects of actions on actions of a common valence:
The second most important predictors of ``m$-$'', ``m$+$'' and ``v$-$''
are ``v$-$'', ``v$+$'' and ``m$-$'', respectively.
The variable ``v$+$'' (verbal positive) represents an exception to this pattern.
However, many of the actions that fall into this category are bilateral
negotiations and diplomatic resolutions that often occur as a result of
diplomatic disputes that are in the ``verbal negative'' category.
The second four columns of $\bl B_3$ represent the
reciprocal effects of actions from $i_2$ to $i_1$ on future
actions from $i_1$ to $i_2$. Similar to the direct effects,
the largest coefficients for three of the four action types
are along the main diagonal. The exception is the ``m$+$''
category (material positive), for which
the 95\% posterior confidence interval contained zero. This reflects
the fact that this category is largely comprised of
actions that involve the provision of economic, military and
humanitarian aid.
Such actions are typically initiated by wealthy countries
with less-developed countries as the target, and so are often
unreciprocated. The final four columns of $\bl B_3$
represent the transitivity effects. While
the results indicate some evidence of transitivity, the magnitude
of such effects is small compared to the direct and reciprocal effects.

%
\begin{table}
\tabcolsep=0pt
\caption{Summary of the posterior distribution of $\bl B_3$}
\label{tab:b3}
\begin{tabular*}{\textwidth}{@{\extracolsep{\fill}}lcccccccccccc@{}}
\hline
& \multicolumn{12}{c@{}}{\textbf{Predictor}} \\[-6pt]
& \multicolumn{12}{c@{}}{\hrulefill} \\
& \multicolumn{4}{c}{\textbf{Direct}} &
\multicolumn{4}{c}{\textbf{Reciprocal}} &
\multicolumn{4}{c@{}}{\textbf{Transitive}} \\[-6pt]
& \multicolumn{4}{c}{\hrulefill} &
\multicolumn{4}{c}{\hrulefill} &
\multicolumn{4}{c@{}}{\hrulefill} \\
\textbf{Outcome}&
\textbf{m}$\bolds{-}$ & \textbf{m}$\bolds{+}$ &
\textbf{v}$\bolds{-}$ & \textbf{v}$\bolds{+}$ &
\textbf{m}$\bolds{-}$ & \textbf{m}$\bolds{+}$ &
\textbf{v}$\bolds{-}$ & \textbf{v}$\bolds{+}$ &
\textbf{m}$\bolds{-}$ & \textbf{m}$\bolds{+}$ &
\textbf{v}$\bolds{-}$ & \textbf{v}$\bolds{+}$ \\
\hline
m$-$ & 0.68 & 0.04 & 0.17 & & 0.20 & 0.02 & 0.12 & 0.02 & 0.02 & & & \\
m$+$ & 0.09 & 0.50 & 0.04 & 0.13 & 0.04 & & 0.02 & & & & 0.04 & \\
v$-$& 0.18 & & 0.61 & 0.12 & 0.13 & 0.03 & 0.21 & & & 0.01 & 0.02 & \\
v$+$& 0.05 & 0.03 & 0.08 & 0.67 & 0.05 & 0.02 & 0.03 & 0.32 & & 0.02 &
0.02 & \\
\hline
\end{tabular*}
\end{table}

The matrix $\bl B_4$ consists of two coefficients representing the
multiplicative effects of one-week lagged
data as compared to one-month lagged data.
Both coefficients of $\bl B_4$ were
positive in every iteration of the Gibbs sampler,
and the posterior distribution of the ratio of the former coefficient
to the latter had a mean of 1.98 and a 95\% posterior confidence interval
of (1.94, 2.03), indicating that the effect of the one-week lagged data was
roughly twice that of the one-month lagged data.

\section{Discussion}\label{sec5}
This article has developed a general
multilinear tensor regression (MLTR) model
for regressing a tensor of correlated outcome data on a tensor of
explanatory variables.
The regression coefficients in such a model are multiplicative in the
parameters, rather than additive as in the more standard class of
linear regression models. As was shown in an example analysis
of longitudinal relational data, in some cases a multiplicative effects
model provides a better representation of the data than a
comparable but more standard
additive effects model.
Additionally, it was shown how the
MLTR model can be extended to estimate a variety of
network effects,
such as reciprocity and transitivity,
as well as temporal effects of lagged data beyond those in a
first-order autoregressive model.

Application of this MLTR model to longitudinal international
relations data provided a quantification of how the
relations and actions of a given country are dependent upon those
of other countries.
Specifically, the application identified those countries
whose actions are predictive of a given country's future actions,
and quantified this predictive dependency.
The strongest dependencies are generally
between countries that are geographically close, with
exceptions being the dependence between
Australia and the United Kingdom, and between the United
States
and several countries.
Furthermore, this application
identified dependencies between different types of
relations and the extent to which these
relations are reciprocated.
In summary, the results of the application indicate that
the relations between a given pair of
countries are dependent on those of other country pairs, and
that data analyses that ignore this fact present an incomplete
picture of the dynamics of international relations.

Like any regression model,
the multilinear tensor regression model could be extended or modified in
many different ways. Of particular use would be an extension to accommodate
data that is binary, ordinal or generally of a form
for which a least squares criteria or
normal error model would be inappropriate.
One possible approach for doing this would be via various link functions,
as is done with generalized linear models.
An alternative approach would be to use a semiparametric
transformation model via a rank likelihood [\citet
{pettitt1982,hoff2007a}], in
which the observed data are modeled as being a nondecreasing function
of a
latent tensor that follows a normal multilinear tensor regression model.
However, for some data types, such as if $\bl Y$ were a sparse binary tensor,
there might not be enough information in the data to provide stable
parameter estimates. Even though the MLTR model
with $\Exp{\bl y} = (\bl B_K \otimes\cdots\otimes\bl B_1 ) \bl x$
constitutes a great simplification as compared to a full model
$\Exp{\bl y} = \bs\Theta\bl x$, 
the MLTR model still has a large number of parameters.
One possible remedy in cases with limited data information is to use
sparsity-inducing penalties, such as $L_1$ penalties on the $\bl B_k$'s.
This would have to be done with some care, as the overall scale
of each $\bl B_k$ matrix is not identifiable.

\begin{appendix}\label{app}

\section{Proofs}\label{appa}

\begin{pf*}{Proof of Proposition~\ref{prop:ptrue1}}
Let $(\tilde{\bl A}, \tilde{\bl B})$ be a pseudotrue parameter
for $(\bl A, \bl B)$. If $\tilde{\bl B} = \bl0$, then
setting $\tilde a_{i,j}$, the $i,j$th element of $\tilde{\bl A}$, to
zero does not change the asymptotic criterion function, and so
$\tilde a_{i,j}= 0$ is a pseudotrue value.
If $\tilde{\bl B}\neq\bl0$, then $\Exp{ \bl X \tilde{\bl B}^T\tilde
{\bl B} \bl X^T}$
is invertible
(assuming, e.g.,
the distribution of $\bl X$
has full support on $\mathbb R^{p_1\times p_2}$), and the
pseudotrue parameter $\tilde{\bl A}$ will satisfy
%
%
\begin{equation}
\tilde{\bl A} = {\mathrm{E}}\bigl[\bl Y \tilde{\bl B} \bl X^T \bigr]
{\mathrm{E}}\bigl[ \bl X \tilde{\bl B}^T\tilde{\bl B} \bl
X^T \bigr]^{-1}. \label{eq:aptrue}
\end{equation}
%
Let $\bl y_i$ and $\bl x_j$ be rows $i$ and $j$ of
$\bl Y$ and $\bl X$, respectively.
If $\bl x_j$ is mean zero and independent of the other rows of $\bl X$,
so that $\Exp{ \bl x_j \bl x_k^T} = \bl0$, then
the $i,j$th element of $\tilde{\bl A} $ is given by
\[
\tilde a_{i,j} = {\mathrm{E}}\bigl[ \bl y_i^T
\tilde{\bl B} \bl x_j \bigr] / {\mathrm{E}}\bigl[ \bl
x_j^T\tilde{\bl B}^T \tilde{\bl B}\bl
x_j \bigr].
\]
If $\bl y_i$ is uncorrelated with $\bl x_j$, then the numerator
and the coefficient are zero.
\end{pf*}

\begin{pf*}{Proof of Proposition~\ref{prop:ptrue2}}
As in the proof of Proposition~\ref{prop:ptrue1},
if\break $\Exp{ \bl X \tilde{\bl B}^T\tilde{\bl B} \bl X^T}$ is invertible,
then the pseudotrue parameter is given by
$\tilde{\bl A}$ in (\ref{eq:aptrue}). Under the assumption
that $\Exp{ \bl x \bl x^T } = \Omega\otimes\Psi$,
we have $\Exp{\bl X \tilde{\bl B}^T\tilde{\bl B} \bl X^T}
= c \Psi$ with $c=\tr(\Omega\bl B^T \bl B)$, and
\begin{eqnarray*}
{\mathrm{E}}\bigl[ \bl Y \bl B \bl X^T \bigr] &= \bigl(
\bl1_{m_2}^T \otimes\bl I_{m_1} \bigr) \bigl[
\Sigma_{yx} \circ \bigl(\bl B \otimes\bl1 \bl1^T \bigr)
\bigr] (\bl1_{p_2} \otimes\bl I_{p_1} ),
\end{eqnarray*}
where ``$\circ$'' is the Hadamard (elementwise) product.
Under the assumption of the proposition,
$\Sigma_{yx} =
\Exp{\bl y \bl x^T} = \Exp{ \Exp{\bl y|\bl x} \bl x^T }
= \bs\Theta\Exp{ \bl x \bl x^T}
= \bs\Theta(\Omega\otimes\bl\Psi)$, which can be expressed
as
\begin{eqnarray*}
&&\pmatrix{ \bs\Theta_{1,1} & \cdots& \bs
\Theta_{1,p_2}
\cr
\vdots& & \vdots
\cr
\bs\Theta_{m_2,1} & \cdots& \bs\Theta_{m_2,p_2} } \pmatrix{
\omega_{1,1} \Psi& \cdots& \omega_{1,p_2} \Psi
\cr
\vdots& &
\vdots
\cr
\omega_{p_2,1}\Psi& \cdots& \omega_{p_2,p_2} \Psi} \\
&&\qquad= \pmatrix{
\displaystyle\sum_{1\leq j_2\leq p_2} \omega_{j_2,1} \bs
\Theta_{1,j_2} \Psi& \cdots& \displaystyle\sum\omega_{j_2,p_2} \bs
\Theta_{1,j_2} \Psi
\cr
\vdots& & \vdots
\cr
\displaystyle\sum
\omega_{j_2,1} \bs\Theta_{m_2,j_2} \Psi& \cdots& \displaystyle\sum
\omega_{j_2,p_2} \bs\Theta_{m_2,j_2} \Psi},
\end{eqnarray*}
where
$\bs\Theta_{i_2,j_2}$ is the $m_1 \times p_1$ matrix describing
the effects of the column $j_2$ of $\bl X$ on column $i_2$ of $\bl Y$.
The expectation $\Exp{\bl Y \bl B \bl X^T}$ is
obtained by multiplying each block of the form
$\sum_{j_2=1}^{p_2} \omega_{j_2,j_2'} \bs\Theta_{i_2, j_2} \Psi$ by
element $i_2,j'_2$ of $\bl B$, and summing the blocks. This results in
an $m_1\times p_1$ matrix given by
\[
{\mathrm{E}}\bigl[\bl Y \bl B \bl X^T \bigr] = \Biggl( \sum
_{i_2=1}^{m_2} \sum_{j_2=1}^{p_2}
\Biggl( \sum_{j_2'=1}^{p_2}
\omega_{j_2,j_2'} b_{i_2, j'_2} \Biggr) \bs\Theta_{i_2,j_2} \Biggr)
\Psi.
\]
Multiplying by the inverse of
$\Exp{\bl X \tilde{\bl B}^T\tilde{\bl B} \bl X^T}$ on the left gives the
pseudotrue parameter $\bl A$ as
\[
\tilde{\bl A} = c^{-1} \sum_{i_2=1}^{m_2}
\sum_{j_2=1}^{p_2} \Biggl( \sum
_{j_2'=1}^{p_2} \omega_{j_2,j_2'} b_{i_2, j_2'}
\Biggr) \bs\Theta_{i_2,j_2}.
\]
The effects of the $j$th row of $\bl X$ on the $i$th row of $\bl Y$
consist of the $i,j$th elements of the $\bs\Theta_{i_2,j_2}$'s.
These are all zero under the assumption of the proposition,
and thus so is $\tilde a_{i,j}$.
\end{pf*}


\section{Details of the MCMC algorithm}\label{appb}
The posterior distribution described in Section~\ref{sec4.3} was
approximated with four separate Gibbs samplers:
three with random starting values and one starting at the least squares
estimates. Each sampler was run for 5500 iterations,
allowing for 500 iterations for convergence to the stationary distribution.
The sampler that started at the least squares estimates appeared to converge
essentially immediately,
whereas the samplers with random starting values appeared to take
between about 50 and 250 iterations to arrive at the
same part of the parameter space.
Recalling that the separate magnitudes of the
$\bl B_k$'s (and the $\Sigma_k$'s) are not
separately identifiable [as $\bl F \otimes\bl G = (c\bl F)\otimes(\bl
G/c)]$, we saved normalized versions
of these parameters from the MCMC output.
The normalization
maintained a constant relative magnitude among $\Vert\bl B_1\Vert^2,\break 
\Vert\bl B_2\Vert^2, \Vert\bl B_3\Vert^2
\Vert\bl B_4\Vert^2$,
but leaves the magnitude of $\bl B_4 \otimes\bl B_3 \otimes\bl B_2
\otimes\bl B_1$
unchanged as compared to doing no normalization.
The $\Sigma_k$'s were rescaled similarly. Further details on this
post-processing of the MCMC output is available from the replication
code available at the author's website.
%
%
\begin{figure}
\includegraphics{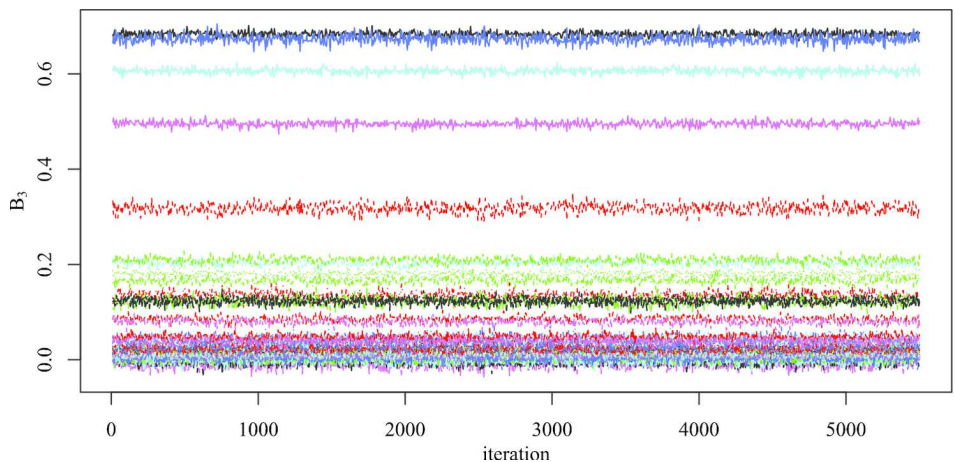}
\caption{Values of the 48 entries of
$\bl B_3$ simulated from the Gibbs sampler,
using the least squares estimates as starting values.}
\label{fig:mcmcdiag}
\end{figure}
Mixing of the Gibbs sampler was very good: Figure~\ref{fig:mcmcdiag}
shows traceplots of the elements of $\bl B_3$, the coefficients
describing the effects of the different action types,
from the Gibbs sampler starting at the least squares estimates.
After convergence, traceplots from the other Gibbs samplers looked
nearly identical.
For example,
the across-sampler standard deviation of the
four posterior mean estimates
was not more than 0.0011 for any element
of any of the $\bl B_k$'s.

\end{appendix}

\section*{Acknowledgments} Replication code for the results in
Section~\ref{sec4} is available at the author's website:
\url{http://www.stat.washington.edu/\textasciitilde pdhoff}.
The author thanks Michael Ward for
guidance with the data.\vadjust{\eject}

%





\printaddresses
\end{document}